\documentclass[12pt]{article}
\usepackage{amssymb,graphicx}
\usepackage{epstopdf}
\usepackage{amsmath,amsfonts}
\usepackage{epsfig}\language = 1
\usepackage{graphicx}

\newcommand{\be}{\begin{equation}}
\newcommand{\ee}{\end{equation}}
\newcommand{\bea}{\begin{eqnarray}}
\newcommand{\eea}{\end{eqnarray}}
\newcommand{\bg}{\begin{gather}}
\newcommand{\eg}{\end{gather}}
\newcommand{\bseq}{\begin{subequations}}
\newcommand{\eseq}{\end{subequations}}

\renewcommand{\ln}{\mathop{\rm ln}\nolimits}

\newcommand{\pd}[0]{\partial}
\newcommand{\al}[0]{\alpha}

\usepackage[cp1251]{inputenc}

\begin{document}
\begin{center}
  {\LARGE \bf Scalar tilt from broken conformal invariance} \\
\vspace{20pt}
M.~Osipov, V.~Rubakov
\vspace{15pt}

\textit{
Institute for Nuclear Research of
         the Russian Academy of Sciences,\\  60th October Anniversary
  Prospect, 7a, 117312 Moscow, Russia}\\

    \end{center}
    \vspace{5pt}

\begin{abstract}
Within recently proposed scenario which explains 
flatness of the spectrum of scalar cosmological
perturbations by a combination of conformal and global
symmetries, we discuss the effect of weak breaking of
conformal invariance. We find that the scalar  power
spectrum obtains a small tilt which depends on
both the strength of conformal symmetry breaking
and the law of evolution of the scale factor.

\end{abstract}

\section{Introduction and summary}

Primordial scalar perturbations in the Universe are 
Gaussian (or nearly Gaussian) and have nearly flat power
spectrum~\cite{Komatsu:2010fb}. 
These properties are nicely obtained in
inflationary theory~\cite{Mukhanov:1981xt}. 
Inflation is not unique
in this respect, 
however~\cite{finelli,minus-bis,Wands:1998yp,Mukohyama:2009gg,Rubakov:2009np,Creminelli:2010ba, Lehners}. 
In particular, nearly Gaussian scalar perturbations with
flat power spectrum may be generated~\cite{Rubakov:2009np} in a theory
of conformal scalar field with negative quartic potential
possessing some global symmetry (e.g., $U(1)$). If conformal invariance is
exact, the latter mechanism is insensitive to the regime
of the cosmological evolution. There is a qualification, though,
that there should exist long enough epoch preceding the hot Big Bang
stage
(so that the horizon problem is solved, at least formally);
the mechanism is assumed to operate at that epoch. Once conformal
invariance is slightly broken (which may be quite natural),
one expects that the spectrum is slightly tilted and depends on
the evolution of the scale factor. The purpose of this note is to
demonstrate the latter properties explicitly.

In Section~\ref{2} we introduce our model and discuss in general terms
the mechanism of the generation of scalar perturbations. 
We then specify the evolution of the scale factor and find
the homogeneous background solution describing the scalar field
rolling down its potential. The power spectrum is calculated
in Section~\ref{sec:Pert}. We indeed find that the spectrum is tilted, and 
that the tilt depends on both the strength of the 
violation of conformal invariance
and the evolution of the scale factor at the epoch when the scalar
field rolls down.

\section{Model and background solution}
\label{2}

The model we consider in this note 
involves a complex scalar field $\phi$
conformally coupled to gravity. 
Its dynamics at the epoch of interest is governed by the
following
action:
\be
S[\phi] = 
\int d^4x \sqrt{-g} \left[ g^{\mu\nu}\pd_\mu\phi^* \pd_\nu\phi 
+ \frac{R}{6} \phi^* \phi - V(\phi) \right] \; .
\label{jul8-10}
\ee
It possesses global $U(1)$ symmetry, and we assume approximate
conformal symmetry. To this end, we choose the following scalar
potential,
\be
\label{jul9-10}
V (\phi)= - h^2 |\phi|^{4+\al}   \; ,
\ee
where $h$ is the coupling constant and the parameter
$\al$ characterizes the deviation from conformal invariance.
The scalar potential is negative, so the homogeneous background field
rolls it down. Without loss of generality, the background
field $\phi_c (t)$ can be chosen real.

The mechanism of the generation of scalar perturbations is as 
follows~~\cite{Rubakov:2009np}.
One assumes that the Universe is homogeneous and isotropic at
the rolling stage and  the
cosmological evolution is dominated by some
other matter
during  that stage and somewhat later.
For $\al = 0$, the dynamics of the
field $\chi = a\cdot \phi = \chi_1 + i\chi_2$ is independent of
the evolution of the scale factor. In that case, the perturbations
of the imaginary part $\chi_2$ behave at early times
as free massless scalar field in Minkowski space. Later on, when
the background $\chi_c$ becomes large, the perturbations of the
phase $\delta \theta = \delta \chi_2/\chi_c$ freeze out.
For small $h$ and $\al =0$ these perturbations are nearly Gaussian
and have flat power spectrum. 

If the potential $V(\phi)$ actually has a minimum at some
large field value,
and the field $\phi$ interacts with matter, the rolling stage
terminates at some late time, and the modulus $\phi_1$ relaxes to the minimum.
Provided that the energy density
of the field $\phi$ is still negligible at that time,
the perturbations of the modulus $\delta \phi_1$ are irrelevant
for small $h$;
what remains are perturbations of the phase.
The phase is no longer conformal scalar field; instead, it minimally
couples to gravity, as any other 
Nambu--Goldstone field~\cite{Voloshin:1982eb}. 
The simplest possibility
is to assume that modes of cosmologically relevant conformal
momenta are superhorizon by the time the modulus relaxes, 
then the perturbations of the
phase remain frozen out\footnote{An implicit assumption here
is that there are no  superhorizon modes that
grow in time. This is indeed the case in the majority of viable
cosmological models.}.

The final part of the scenario is reprocessing the perturbations of
the phase 
into adiabatic perturbations. There are at least two mechanisms
for that. The phase can actually be a {\it pseudo-}Nambu--Goldstone field
of relatively small mass, and serve as 
curvaton~\cite{Linde:1996gt,Dimopoulos:2003az}.
An alternative is that the phase field $\theta$ interacts with matter 
in such a way that the masses and/or widths of some heavy particles
depend on $\theta$ (say, $M= M_0 + \epsilon \theta$ and/or
$\Gamma = \Gamma_0 + \epsilon \theta$).
If the latter particles, when non-relativistic, temporarily dominate 
the cosmological expansion, inhomogeneities in $\theta$, and hence in
$M$ and/or $\Gamma$ 
induce adiabatic perturbations 
$\zeta \sim \delta M/M, \delta \Gamma/\Gamma
$~\cite{Dvali:2003em,Vernizzi:2003vs}. 
In either case\footnote{Note that both the pseudo-Nambu--Goldstone
curvaton mechanism and modulated decay mechanism require
explicit breaking of global $U(1)$.}, to the linear order one has
\be
   \zeta \propto \delta \theta \; ,
\label{jul9-20}
\ee
so the adiabatic perturbations have flat power spectrum in the
model with $\al =0$.

In this note we are interested in the case of slightly broken
conformal invariance, i.e., $\al \neq 0$, but 
\[
|\al| \ll 1 \;.
\]
In that case, the scale factor does not drop out from the field
equation, and we have to specify the cosmological model.
We choose for definiteness contracting Universe filled with matter
with stiff equation
of state $w>1$, e.g., some version of the ekpyrotic/bouncing Universe 
scenario~\cite{Khoury:2001wf,minus-bis,Lehners}. In terms of conformal time,
the evolution of the scale factor is
\be
a(\eta) = {\mathcal A}(-\eta)^{p} \; ,
\label{jul8-2}
\ee
where 
\[
p = \frac{2}{1+3w} \;, \;\;\;\; {\mathcal A}= \mbox{const} \; .
\]
We begin with the homogeneous background solution that rolls down the
scalar potential. 

It is convenient to perform the following
change of variables:
\begin{align}
\phi &= a^{-\frac{6\al}{6+\al}} \Psi \; ,
\nonumber\\
d\xi &= a^{-\beta} d\eta \; ,
\label{jul8-3}
\end{align}
where 
\[
\beta = \frac{2\alpha}{6+\alpha}\;.
\]
The motivation for this change of variables is to 
get rid of the first time derivative in the field
equation and simultaneously
remove 
the time variable from the scalar potential. 
The action in new variables is
\begin{align}
S[\Psi] = \int d\xi d^3 x  &\left\{ |\Psi'|^2 
- a^{2\beta} |\partial_i \Psi|^2  
+ h^2 |\Psi|^{4+\al}  \phantom{\frac{1}{2}} \right.
\nonumber\\ 
& \left.+ 
\left[ 
\frac{\al(6+2\al)}{(6+\al)^2} 
\left( \frac{a'}{a}\right)^2  -\frac{\al}{6+\al}\frac{a''}{a}
\right] \cdot  |\Psi|^2  \right\} \; ,
\label{jul8-1}
\end{align}
where 
prime denotes the derivative with respect to $\xi$, and the 
second term vanishes for the homogeneous background solution.
As the field rolls down its potential towards large values,
the last term in (\ref{jul8-1}) becomes subdominant compared to
the third term. Hence, at late rolling stage the equation for
the background reads
\begin{equation}
\label{Bckgr}
\Psi_c'' - (2+\al/2)h^2\cdot\Psi_c^{3+\al} = 0 \; ,
\end{equation}
where we assume that the background field $\Psi_c$ is real.
The late time attractor solution to this equation is
\be
\Psi_c (\xi) = \left[h(1+{\al}/{2})(\xi_{*} - 
\xi) \right]^{-\frac{2}{2+\al}}\; ,
\label{BckgrSol}
\ee
where $\xi_*$ is the intergation constant. We assume in what follows
that 
\be 
\xi_* <0 \; ,
\label{jul8-5}
\ee
so that the entire rolling epoch occurs at the contracting stage.

We have checked numerically that homogeneous solutions
to the complete field equation derived from the  action
(\ref{jul8-1}) indeed approach the asymptotics
(\ref{BckgrSol}) as the field rolls down
its potential.

\section{Scalar perturbations}
\label{sec:Pert} 

Our main purpose 
is to study linear
perturbations of the field $\phi$ in the background 
(\ref{jul8-2}), (\ref{BckgrSol}). 
In line with the above discussion, we concentrate on the 
perturbations of the imaginary part.
In terms of the variables (\ref{jul8-3}), they obey the
following equation,
\begin{equation}
\label{Perturbations}
\delta\Psi_2'' + k^2 a^{2\beta}\cdot\delta\Psi_2 - 
(2+\al/2)h^2\Psi_c^{2+\al}\cdot \delta\Psi_2 = 0 \; ,
\end{equation}
where we again neglected the contribution that  comes from
the last term in the action (\ref{jul8-1}). Indeed, the latter
contribution is of order $\delta \Psi_2 \cdot \al/\xi^2$. In view of
(\ref{jul8-5}),
this is small
compared to the last term in the left hand side of Eq.~(\ref{Perturbations}),
which is of order
 $\delta \Psi_2\cdot 1/(\xi_* - \xi)^2$.

For the same reason, there is 
the inequality
\[
   \beta \frac{a^\prime}{a} \ll \frac{\Psi_c^\prime}{\Psi_c} \; .
\]
This inequality means that the function $a^{2\beta}$ slowly
varies in time as compared to $\Psi_c^2$.
This suggests the following approach for obtaining the solution to
Eq.~(\ref{Perturbations}). At early times, when
\be
k^2 a^{\beta} \gg h^2\Psi_c^{2+\al} \; ,
\label{jul8-6}
\ee
one neglects the last term in the left hand side of Eq.~(\ref{Perturbations})
and
makes use of the WKB approximation. Note that the inequality 
(\ref{jul8-6}) can be considered as the condition that the mode
of conformal momentum $k$ is in sub-''horizon'' regime,
the effective ``horizon''  being due to the
evolution of the background field $\Psi_c$. 
At  
late times, when the inverse
inequality holds, one neglects the second term in
Eq.~(\ref{Perturbations}).
To match the two solutions, one solves Eq.~(\ref{Perturbations}) at
intermediate times around the  ``horizon'' exit, using the approximation of the
time-independent scale factor.
In fact, the last two steps can be combined: since the term involving the
scale factor is irrelevant at late times anyway, the solution to the equation
with the
time-independent scale factor is valid both at intermediate
and late times.

The WKB solution in the sub-''horizon'' regime (\ref{jul8-6}) is
\be
\delta\Psi_2 = \frac{1}{(2\pi)^{3/2}
\sqrt{2k a^{\beta}}}\cdot 
\exp\left(i k \int a^{\beta} d\xi \right)  \hat{A}_{\bf k}^{\dagger} + h.c. \; ,
\label{jul8-11}
\ee
where $\hat{A}_{\bf k}^\dagger$ and  $\hat{A}_{\bf k}$ are creation and 
annihilation operators obeying the standard commutational relations.
This solution is equivalent to
\be
\delta \chi_2 \equiv \delta (a\phi_2)= \frac{1}{(2\pi)^{3/2}
\sqrt{2k}}\cdot 
\mbox{e}^{ik \eta}  \hat{A}_{\bf k}^{\dagger} + h.c.
\label{jul8-7}
\ee
This is the expected result, since by neglecting the scalar
potential in the sub-''horizon'' regime,
we are dealing with the theory of massless scalar
field conformally coupled to gravity, whose modes are given
precisely by (\ref{jul8-7}).

At the time around the  ``horizon'' exit and later
we neglect the dependence
of $a^\beta$ on time and
set
\[
a^\beta (\xi) =  a^\beta (\xi_\times) \equiv a_\times^\beta \; ,
\]
where the ``horizon'' crossing time $\xi_\times$ is found
from the relation
 \[
k^2 a_\times^{2\beta} = h^2\Psi_c^{2+\al} (\xi_\times) \; .
\]
Then Eq.~(\ref{Perturbations}) becomes
\begin{equation}
\label{aFixed}
\delta\Psi_2'' + 
k^2 a_\times^{2\beta}\cdot\delta\Psi_2 - 
\frac{2+\al/2}{(1+\al/2)^2} \cdot \frac{1}{(\xi_* - \xi)^2}\delta\Psi_2 = 0.
\end{equation}
Its solution that matches (\ref{jul8-11}) at $ka_\times^\beta (\xi_*-\xi) \gg
1$ is
\begin{equation}
\label{Interm}
\delta\Psi_2 = \frac{1}{4\pi}
\sqrt{\frac{\xi_* - \xi}{2}}\cdot H_\nu^{(1)} 
\left[k a_\times^{\beta}(\xi_* - \xi) \right] \hat{A}_{\bf k}^{\dagger} + h.c. \; ,
\end{equation}
where the index of the
Hankel function equals
\[
\nu = \frac{1}{2}\cdot\frac{6+\al}{2+\al} \; .
\]
We find from (\ref{Interm}) that the late time asymptotics of the perturbations
of the phase is given by
\be
\delta \theta \equiv \frac{\delta \Psi_2}{\Psi_c}
= \frac{1}{4\pi^{3/2}} \frac{h^{\frac{2}{2+\alpha}}}{k^\nu a_\times^{\nu \beta}}
 \hat{A}_{\bf k}^{\dagger} + h.c.\; ,
\label{jul9-2}
\ee
were we neglected correction of order $\alpha$ in the numerical
pre-factor.
As explained in Ref.~\cite{Rubakov:2009np}, the fact that
the phase freezes out at late times is due to the global
$U(1)$ symmetry of the action~(\ref{jul8-10}).

To proceed further, we have to find the ``horizon'' exit time $\xi_\times$.
It is determined by the relation
\be
k^2 a_\times^{\beta} 
= h^2 \Psi_c^{\al+2} (\xi_\times)\approx \frac{1}{(\xi_* - \xi_\times)^2}
\label{jul9-1}
\ee
We proceed under the assumption that the relevant scales are
superhorizon in conventional sense by the end of the rolling stage.
This implies, in particular, that 
\[
    k|\eta_*| \ll 1 \; ,
\]
where $\eta_*$ is the moment of conformal time corresponding to $\xi = \xi_*$.
The latter inequality implies $|\xi_\times| \gg |\xi_*|$, so we obtain
from (\ref{jul9-1}) that
\[
 k a_\times^{\beta}=|\xi_\times|^{-1} = k^{1-p\beta}{\mathcal A}^{\beta} \; ,
\]
where we keep the terms of the first order in $\alpha$ in the exponent.
Finally, according to (\ref{jul9-2}), the power spectrum of the perturbations
of the phase is
\[
{\mathcal P}(k) = \frac{h^2}{(2\pi)^2} 
 ({\mathcal A} h)^{-\al} 
\cdot k^{\al(1+p)} \; .
\]
In accord with our expectation, 
violation of conformal invariance leads to the tilted power spectrum.

As discussed in Section~\ref{2}, perturbations of the phase are assumed
to be
reprocessed into adiabatic scalar perturbations at later epoch
of the cosmological evolution, see (\ref{jul9-20}). 
The adiabatic perturbations inherit
the tilt of the spectrum, so that in our scenario
\[
n_s - 1 = \al(1+p) \; . 
\]
Note that the tilt depends not only on the 
conformal symmetry breaking parameter $\al$ but also on the evolution
of the scale factor parametrized by the exponent $p$ in our case.

We conclude by noticing that breaking of coformal  invariance
may result,
instead of the power-law potential
(\ref{jul9-10}), in the logarithmic effective potential
\[
V(\phi) = - g^2 \ln \frac{\phi}{\mu} \; ,
\]
where   $\mu$ is assumed to be small.
This possibility is, in fact, quite natural from the
viewpoint of quantum field theory. The 
logarithmic potential can be obtained by taking the formal
limit $\alpha \to 0$, $\alpha h^2 = \mbox{const}$.
The resulting power spectrum also has logarithmic behavior,
\[
{\mathcal P}(k) = \frac{g^2}{(2\pi)^2} 
(1+p) \ln \frac{k}{k_*} \; .
\]
where $k_* \sim \mu g (a_\times k^p)$ is almost independent of
$k$, and $k/k_*$ is large for small $\mu$. Hence, logarithmic
violation of conformal invariance implies very mild deviation of the
scalar power spectrum from the Harrison--Zeldovich form.

The authors are indebted to A.~Barvinsky, S.~Dubovsky,
E.~Komatsu, M.~Libanov, S.~Ramazanov and P.~Tinyakov for helpful discussions.
V.R. is grateful to the organizers of the Yukawa International
Seminar ``Gravity and Cosmology 2010'', where part of this work has
been done, for hospitality.
 This work
has been supported in part by Russian Foundation for Basic
Research grant 08-02-00473.


\end{document}